\documentclass[aip, reprint, physrev]{revtex4-2}

\usepackage{graphicx}
\usepackage{dcolumn}
\usepackage{bm}
\usepackage[utf8]{inputenc}
\usepackage[T1]{fontenc}
\usepackage{amsmath}
\usepackage{mathptmx}
\usepackage{etoolbox}
\usepackage{accents}
\usepackage{colortbl}
\usepackage{enumitem}
\usepackage{comment}
\usepackage{booktabs}
\usepackage{array,multirow}
\usepackage[table,xcdraw]{xcolor}
\usepackage[version=3]{mhchem} 
\usepackage{orcidlink}

\newcommand*{\ghj}[1]{\textcolor{black}{#1}}
\newcommand*{\jmlm}[1]{\textcolor{black}{#1}}

\newcommand{\kcalmol}{\mbox{kcal$\cdot$mol$^{-1}$}}
\newcommand{\kJmol}{\mbox{kJ$\cdot$mol$^{-1}$}}

\makeatletter
\def\@email#1#2{
 \endgroup
 \patchcmd{\titleblock@produce}
  {\frontmatter@RRAPformat}
  {\frontmatter@RRAPformat{\produce@RRAP{*#1\href{mailto:#2}{#2}}}\frontmatter@RRAPformat}
  {}{}
}
\makeatother
\begin{document}


\title[CFOUR open-shell CCSDTQ]{A new open-shell CCSDTQ implementation and its application to the basis set convergence of post-CCSDT(Q) corrections in computational thermochemistry}

\author{Aditya Barman\orcidlink{0009-0003-3863-2564}} 
\thanks{Equally contributing authors}
\affiliation{Department of Molecular Chemistry and Materials Science, Weizmann Institute of Science, 7610001 Re\d{h}ovot, Israel}

\author{Gregory H. Jones*\orcidlink{0000-0003-3275-1661}}
\thanks{Equally contributing authors}
\email{ghjones@mit.edu}
\affiliation{Quantum Theory Project, Department of Chemistry,
University of Florida, Gainesville, FL 32611, USA}
\affiliation{Present address: Department of Chemistry, Massachusetts Institute of Technology, Cambridge, MA 02139, USA}

\author{Jan M. L. Martin*\orcidlink{0000-0002-0005-5074}} 
\thanks{On sabbatical at the Quantum Theory Project}
    \email{gershom@weizmann.ac.il}    
    \affiliation{Department of Molecular Chemistry and Materials Science, Weizmann Institute of Science, 7610001 Re\d{h}ovot, Israel}

 \date{CPLETT-26-1042: Revised manuscript \today}

\begin{abstract}
We extend the CCSDTQ implementation in CFOUR to UHF and ROHF references and demonstrate its efficiency. We apply it to basis set convergence of post-CCSDT(Q) corrections for the W4-08 thermochemical dataset. Convergence of (Q)$_\Lambda$--(Q) is relatively rapid. For difficult species (e.g., \ce{B2}, \ce{O3}), CCSDTQ--CCSDT(Q)$_\Lambda$  may converge more slowly than (5)$_\Lambda$, but the effects and and basis-set trends oppose each other. \jmlm{Alternatives to a single-shot CCSDTQ(5)$_\Lambda$--CCSDT(Q)$_\Lambda$ correction are evaluating (5)$_\Lambda$  either in a truncated cc-pVDZ(p,s) basis set, or by means of frozen natural orbitals.} Our best computed adiabatic electron affinity of ozone is in excellent agreement with experiment.
\end{abstract}

\maketitle

\section{Introduction}\label{sec:intro}

In the past two decades, great strides have been made in the area of computational thermochemistry, as reviewed by Karton.\cite{Karton2022} It is well known (e.g.,\citenum{jmlm173,jmlm200,jmlm205})  that the gold standard CCSD(T) method\cite{Rag89,Wat93} actually benefits from an error compensation between neglect of higher-order triples $T_3-(T)$, which are almost always antibonding, and connected quadruples (Q), which are universally bonding. The first genuine step up from CCSD(T) was the CCSDT(Q) method, as introduced by Bomble et al.\cite{mrcc8} in 2004 and by K\'allay and Gauss\cite{KallayGauss2005} in 2005. And indeed, it has pride of place in economical high-accuracy computational thermochemistry schemes, such as W3.2\cite{jmlm173} and its lower-cost variant by Chan and Radom,\cite{ChanRadom} as well as the lower levels of the HEAT (high-accuracy extrapolated ab initio theory\cite{HEAT,HEAT2,HEAT3,HEAT4}) approach of the late lamented Stanton and co-workers. 
 
 However, for higher accuracy levels, such as achievable by the Weizmann-4 (W$4$)\cite{jmlm200,jmlm205} and W5preview\cite{jmlm340} approaches of the Martin group, as well as by HEAT and recently its emerging successor SuperHEAT\cite{Thorpe2021_HEAT,Thorpe2023} variant of Thorpe, Franke et al., still higher-order corrections are required. The basis set convergence of these higher-order corrections has been studied in some detail (e.g., Refs.\cite{jmlm205,jmlm330}). 
 
 Recent studies indicate\cite{jmlm326,jmlm330} that the $\Lambda$ coupled cluster series introduced by Stanton\cite{lambdastanton1,lambdastanton2} and independently by Bartlett\cite{lambdabartlett1,lambdabartlett2}, converge faster and more smoothly to the full CI limit than the conventional CC($n-1$)($n$) series. This is particularly true for CCSDT(Q)$_\Lambda$ and CCSDTQ(5)$_\Lambda$ versus CCSDT(Q) and CCSDTQ(5). It would, however, have been highly desirable to investigate the basis set convergence of such contributions in greater detail. 

 Alas, the very high cost of such calculations using the general coupled cluster engine\cite{KallaySurjan2001,KallayGauss2005,KallayGauss2008} in MRCC\cite{MRCC2025,MRCCcode} imposed a limit in Ref.\citenum{jmlm330}, at least for open-shell systems. For closed-shell systems, a fast CCSDTQ\cite{OliphantAdamowicz1991CCSDTQ,KucharskiBartlett1992CCSDTQ} implementation was already available\cite{NCC2} inside  CFOUR,\cite{CFOUR,CFOURcode} specifically in the NCC  (new coupled cluster\cite{NCC1,NCC2}) module built on the TBLIS tensor contraction library.\cite{TBLIS}
 
One of us (Gregory H. Jones) just extended the CCSDTQ implementation in CFOUR to UHF and ROHF references. We will document this implementation below, as well as apply it to the research question in the title. 
 
 We will show below that $T_4 - (Q)_\Lambda$ does converge comparatively rapidly with the basis set. However, we will also show that the basis set convergence of connected quintuples kind of runs in counter-phase, and that, as a result, the difference between CCSDTQ(5)$_\Lambda$ and CCSDT(Q)$_\Lambda$ is relatively insensitive to the basis set. Nevertheless, an economical combination of CCSDTQ--CCSDT(Q)$_\Lambda$ in a small basis set with $(5)_\Lambda$ in an even smaller one (or in a small subset of frozen natural orbitals, as we described in Ref.\cite{jmlm344}) appears to be valuable in practical situations.
 
\section{Computational Details}\label{sec:methods}

All electronic structure calculations in this work were carried out on the CHEMFARM HPC system at the Weizmann Institute of Science. All calculations 
\jmlm{at the CCSD(T), CCSDT, CCSDT(Q), CCSDT(Q)$_\Lambda$, and CCSDTQ levels,
regardless of whether the reference determinant was RHF, ROHF, or UHF,} were performed using a development version of the CFOUR\cite{CFOUR} electronic structure program system. The higher-order CCSDTQ(5)$_\Lambda$ calculations have been carried out using the arbitrary-order coupled cluster code\cite{KallaySurjan2001,KallayGauss2005,KallayGauss2008} in MRCC\cite{MRCC2025,MRCCcode} as developed by the K\'allay group. 

Two families of basis sets were considered. The first are the Dunning correlation consistent\cite{Dunning1989,Woon1993} polarized double, triple, and quadruple-zeta basis sets (cc-pVDZ, cc-pVTZ,  and cc-pVQZ), as well as variants where the top angular momenta were truncated. These we denote cc-pVDZ(p,s), cc-pVTZ(d,p), and cc-pVQZ(f,d) respectively. 
The second family are the double and triple-zeta ANO (atomic natural orbital\cite{Almlof1987}) basis sets, ano-pVDZ and ano-pVTZ, of Neese and Valeev.\cite{Neese2011b} We surmised that these atomic natural orbital basis sets converge somewhat faster to the one-particle basis set limit for these contributions.

As in Ref.\cite{jmlm340}, the sample considered is the W4-08 thermochemical benchmark\cite{jmlm215}. These 96 diatomics and small polyatomics cover a wide range of static correlation regimes (from purely dynamical to strong static correlation) and include both closed- and open-shell molecules, as well as both first-row and second-row compounds.

 Reference geometries were optimized at the all-electron CCSD(T) level\cite{Rag89} with a cc-pwCVQZ basis set.\cite{Peterson2002} (The 1s deep core orbital on elements Al--Cl was however frozen, as ordinary core-valence basis sets are unsuitable for deep-core correlation anyway.) These geometries were taken verbatim from Barman et al.,\cite{jmlm340} and used as-is without further reoptimization.  For the open-shell species, ROCCSD(T) optimizations were carried out there. 

For the ozone electron affinity, some CCSDTQ and CCSDTQ(5)$_\Lambda$ calculations were carried out in Papajak-Truhlar calendar basis sets,\cite{PapajakTruhlar} specifically, jun-cc-pV$n$Z ($n$=D,T) which corresponds to omitting the diffuse function with the highest angular momentum on nonhydrogen atoms (and all diffuse functions on hydrogen, which is immaterial for ozone).

\subsection{Implementation details of the open-shell CCSDTQ capabilities}
The NCC coupled-cluster module of the CFOUR program package was extended to cover open-shell CCSDTQ, based on the TBLIS framework\cite{TBLIS}. The implementation makes no assumptions about the nature of the single determinant reference function (i.e. the Fock operator is not assumed to be diagonal, nor are the $f_{ia}$ terms assumed to be zero). This choice presents a negligible computational overhead, and allows for the use of alternative reference functions such as ROHF and QRHF.
The equations are derived diagrammatically, and the factorization follows the closed-shell CCSDTQ implementation in NCC\cite{NCC1,NCC2}, albeit without the benefit of spin-adaptation. \ghj{The diagrams and their accompanying equations are presented in the Supplementary Material.}
The current implementation holds all amplitudes and intermediates in RAM with the exception of the six-index intermediates, which may have up to 10 unique spin combinations in some cases \ghj{(Figure S1)}. These amplitude-dependent intermediates are generated in batches, contracted with the appropriate amplitudes to generate their contribution to the $T_4$ residuals, then immediately discarded. The minimum batch size is a block indexed by a single combination of occupied indices of the six-index intermediate which are held in common with the $T_4$ residuals.

\ghj{All amplitudes and intermediates are stored in a compact, generalized triangular form taking advantage of the antisymmetry of equivalent indices of the same spin as well as the direct product decomposition approach. Quantities with more than four indices are stored as a list of dense blocks in the above format, indexed by some subset (see SI) of the occupied orbitals, which we will call the ``sparse'' indices. As necessary, input tensors are unpacked to the minimal degree, generating intermediate input tensors which are dense in the contracted-over indices. Similarly, wherever equivalent indices of the same spin in the dense part of a result tensor come from a combination of the two input tensors (i.e. equivalent lines from different vertices), a single contraction is performed into a zeroed, temporary result of minimally-unpacked shape. This temporary tensor is then added antisymmetrically to the fully packed target, thus fully accounting for the antisymmetrizer in these indices using only a single contraction. Antisymmetrization of the ``sparse'' indices of the same spin is handled via a wrapper function which performs a contraction for each necessary index permutation. Antisymmetrization of opposite-spin indices of any type is performed by hand as separate contractions. As this involves the exchange of two indices of opposite spin, these contractions involve different spin cases of the input tensors. With the above strategy, only one contraction per unique combination of the two input and result spin cases must be written, although it is clear not every combination of spin cases is allowed. This greatly reduces the programming effort required, reducing a naive count of 1440 terms in $ P(ab/cd) P(ij/kl) \mathcal{W}^{abn}_{ijf} t^{fcd}_{nkl} $ (6 permutations $\times$ 6 permutations $\times$ 10 $\mathcal{W}$intermediate spin cases $\times$ 4 $T_3$ spin cases) to 26 hand-written function calls.}

\section{Results and Discussion}\label{sec:discussion}

\begin{figure*}[!htbp]
    \centering
    \includegraphics[width=0.5\textwidth]{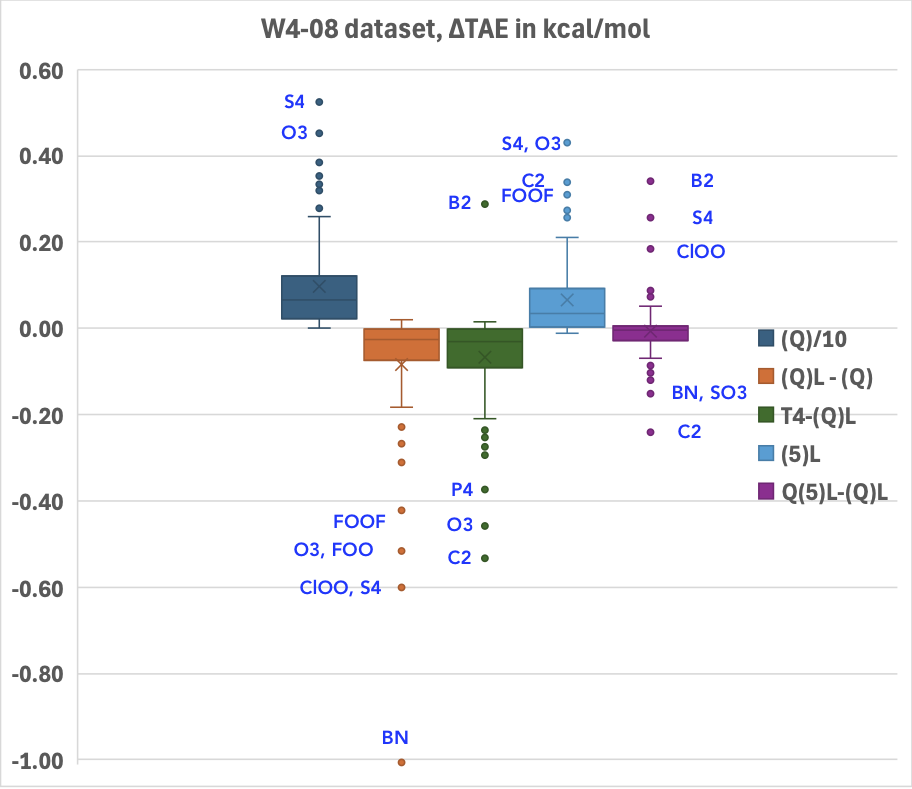}
    \caption{Box-and-whiskers plot of the total atomization energy contributions of higher-order corrections in the W4-08 dataset. (Q) was scaled by 1/10 in order to better accommodate the vertical axis.}
    \label{fig:box-whiskers}
\end{figure*} 

Full numerical results can be found in the Supplementary Material. The various post-CCSDT components with the largest basis set feasible are presented in Table~\ref{tab:components}, while statistics for all basis sets considered can be found in Table~\ref{tab:W4-08}

First, for perspective, let us consider (Q). For accurate work, this clearly needs to be extrapolated from cc-pV\{T,Q\}Z basis sets; presently, we use the extrapolation formula of Karton\cite{Karton2020}, which covers about 0.06 \kcalmol\ RMS from the cc-pVQZ basis set.

\begin{table}[!htbp]
\caption{Post-CCSDT components of the W4-08 total atomization energies (kcal/mol) with the largest basis sets feasible}\label{tab:components}
{
\footnotesize 

\begin{tabular}{clllll}
\hline\hline
          &  &   & \textbf{Best}                       & \textbf{Best}                    & \textbf{Best}\\
 
          & \textbf{V\{T, Q\}Z} &  \textbf{VQZ} & \textbf{available}                       & \textbf{available}                    & \textbf{available}\\
            & \textbf{(Q)} & \textbf{(Q)$_\Lambda$ - (Q)}                       & \textbf{T4-(Q)$_\Lambda$}                    & \textbf{(5)$_\Lambda$} &\textbf{Q(5)$_\Lambda$-(Q)$_\Lambda$}\\
\hline
\ce{B2H6	}	&	\cellcolor[HTML]{FCF7FA	}	0.196	&	\cellcolor[HTML]{F8FAFB	}	0.002	&	\cellcolor[HTML]{FED680	}	0.010$^c$	&	\cellcolor[HTML]{FCF9FC	}	-0.003$^b$	&	\cellcolor[HTML]{FCFAFD	}	0.007$^b$	\\
\ce{BHF2	}	&	\cellcolor[HTML]{FCF3F6	}	0.353	&	\cellcolor[HTML]{F6F9FA	}	-0.010	&	\cellcolor[HTML]{FFE583	}	-0.027$^c$	&	\cellcolor[HTML]{FCFAFD	}	-0.007$^a$	&	\cellcolor[HTML]{DEEFE5	}	-0.042$^a$	\\
\ce{BF3	}	&	\cellcolor[HTML]{FCEFF2	}	0.508	&	\cellcolor[HTML]{F7FAFA	}	-0.005	&	\cellcolor[HTML]{FDEA83	}	-0.049$^c$	&	\cellcolor[HTML]{FCF9FC	}	-0.004$^a$	&	\cellcolor[HTML]{CDE9D6	}	-0.064$^a$	\\
\ce{C2H6	}	&	\cellcolor[HTML]{FCF5F8	}	0.278	&	\cellcolor[HTML]{F9FBFD	}	0.012	&	\cellcolor[HTML]{FFE182	}	-0.018$^c$	&	\cellcolor[HTML]{FCF3F5	}	0.017$^a$	&	\cellcolor[HTML]{FBFBFE	}	-0.003$^a$	\\
\ce{H2CN	}	&	\cellcolor[HTML]{FCEDF0	}	0.599	&	\cellcolor[HTML]{F0F7F5	}	-0.049	&	\cellcolor[HTML]{FFE183	}	-0.019	&	\cellcolor[HTML]{FCEFF2	}	0.032$^b$	&	\cellcolor[HTML]{FCF8FB	}	0.018$^c$	\\
\ce{NCCN	}	&	\cellcolor[HTML]{FBB8BB	}	2.608	&	\cellcolor[HTML]{ECF5F1	}	-0.081	&	\cellcolor[HTML]{B7D67F	}	-0.270$^c$	&	\cellcolor[HTML]{FAABAD	}	0.256$^a$	&	\cellcolor[HTML]{F0F7F4	}	-0.012$^a$	\\
\ce{CH2NH2	}	&	\cellcolor[HTML]{FCF2F5	}	0.389	&	\cellcolor[HTML]{F7FAFB	}	-0.001	&	\cellcolor[HTML]{FFE583	}	-0.026$^c$	&	\cellcolor[HTML]{FCF1F4	}	0.021$^a$	&	\cellcolor[HTML]{F7FAFB	}	-0.007$^a$	\\
\ce{CH3NH	}	&	\cellcolor[HTML]{FCF4F7	}	0.311	&	\cellcolor[HTML]{F8FAFB	}	0.001	&	\cellcolor[HTML]{FFE082	}	-0.015$^c$	&	\cellcolor[HTML]{FCF3F6	}	0.016$^a$	&	\cellcolor[HTML]{FCFCFF	}	0.001$^a$	\\
\ce{CH3NH2	}	&	\cellcolor[HTML]{FCF3F6	}	0.380	&	\cellcolor[HTML]{F9FBFD	}	0.013	&	\cellcolor[HTML]{FFE683	}	-0.029$^c$	&	\cellcolor[HTML]{FCF1F4	}	0.023$^a$	&	\cellcolor[HTML]{F6F9FA	}	-0.009$^a$	\\
\ce{CF2	}	&	\cellcolor[HTML]{FCEBEE	}	0.652	&	\cellcolor[HTML]{F1F7F6	}	-0.041	&	\cellcolor[HTML]{FFE984	}	-0.037$^d$	&	\cellcolor[HTML]{FCFAFD	}	-0.008$^b$	&	\cellcolor[HTML]{D4EBDC	}	-0.051$^b$	\\
\ce{N2H	}	&	\cellcolor[HTML]{FCE4E7	}	0.934	&	\cellcolor[HTML]{EDF6F2	}	-0.07	&	\cellcolor[HTML]{F9E983	}	-0.061$^d$	&	\cellcolor[HTML]{FCE3E6	}	0.068$^b$	&	\cellcolor[HTML]{FCFAFD	}	0.017$^c$	\\
\ce{t-N2H2	}	&	\cellcolor[HTML]{FCE4E7	}	0.944	&	\cellcolor[HTML]{F5F9F9	}	-0.019	&	\cellcolor[HTML]{F4E883	}	-0.076$^d$	&	\cellcolor[HTML]{FCE3E6	}	0.068$^b$	&	\cellcolor[HTML]{F4F9F8	}	0.005$^c$	\\
\ce{N2H4	}	&	\cellcolor[HTML]{FCEFF2	}	0.513	&	\cellcolor[HTML]{F9FBFD	}	0.012	&	\cellcolor[HTML]{FFEA84	}	-0.040$^c$	&	\cellcolor[HTML]{FCF0F3	}	0.027$^b$	&	\cellcolor[HTML]{F0F7F4	}	-0.015$^b$	\\
\ce{FO2	}	&	\cellcolor[HTML]{FAA0A2	}	3.553	&	\cellcolor[HTML]{ACDBBA	}	-0.512	&	\cellcolor[HTML]{CDDC81	}	-0.200$^c$	&	\cellcolor[HTML]{FAA7A9	}	0.269$^a$	&	\cellcolor[HTML]{FCDFE2	}	0.083$^a$	\\
\ce{FOOF	}	&	\cellcolor[HTML]{FA9396	}	4.031	&	\cellcolor[HTML]{B9E1C5	}	-0.421	&	\cellcolor[HTML]{9FCF7E	}	-0.346$^b$	&	\cellcolor[HTML]{FA9B9D	}	0.309$^a$	&	\cellcolor[HTML]{DBEEE2	}	0.000$^a$	\\
\ce{AlF3	}	&	\cellcolor[HTML]{FCEFF2	}	0.503	&	\cellcolor[HTML]{F7FAFA	}	-0.007	&	\cellcolor[HTML]{FCEA83	}	-0.053$^c$	&	\cellcolor[HTML]{FCF8FB	}	-0.001$^a$	&	\cellcolor[HTML]{CCE8D5	}	-0.068$^a$	\\
\ce{Si2H6	}	&	\cellcolor[HTML]{FCF8FB	}	0.187	&	\cellcolor[HTML]{F8FAFB	}	0.001	&	\cellcolor[HTML]{FFD981	}	0.001$^c$	&	\cellcolor[HTML]{FCF7FA	}	0.003$^a$	&	\cellcolor[HTML]{FCFBFE	}	0.003$^a$	\\
\ce{P4	}	&	\cellcolor[HTML]{FAACAF	}	3.071	&	\cellcolor[HTML]{F7FAFA	}	-0.006	&	\cellcolor[HTML]{97CD7E	}	-0.374$^c$	&	\cellcolor[HTML]{FBB8BB	}	0.211$^a$	&	\cellcolor[HTML]{6CC183	}	-0.066$^a$	\\
\ce{SO2	}	&	\cellcolor[HTML]{FBCBCE	}	1.892	&	\cellcolor[HTML]{E7F3ED	}	-0.112	&	\cellcolor[HTML]{C7DA80	}	-0.221$^d$	&	\cellcolor[HTML]{FBD0D3	}	0.132$^b$	&	\cellcolor[HTML]{ADDCBB	}	-0.102$^b$	\\
\ce{SO3	}	&	\cellcolor[HTML]{FBC6C9	}	2.076	&	\cellcolor[HTML]{EDF6F2	}	-0.071	&	\cellcolor[HTML]{C2D980	}	-0.237$^c$	&	\cellcolor[HTML]{FBD3D6	}	0.122$^a$	&	\cellcolor[HTML]{96D3A7	}	-0.147$^a$	\\
\ce{OCS	}	&	\cellcolor[HTML]{FBD4D7	}	1.559	&	\cellcolor[HTML]{F1F7F5	}	-0.046	&	\cellcolor[HTML]{E4E382	}	-0.128$^d$	&	\cellcolor[HTML]{FCE1E4	}	0.074$^b$	&	\cellcolor[HTML]{CCE8D6	}	-0.060$^b$	\\
\ce{CS2	}	&	\cellcolor[HTML]{FBC8CB	}	2.014	&	\cellcolor[HTML]{F0F7F5	}	-0.05	&	\cellcolor[HTML]{E8E482	}	-0.115$^d$	&	\cellcolor[HTML]{FBD7D9	}	0.110$^b$	&	\cellcolor[HTML]{F8FAFB	}	0.035$^c$	\\
\ce{S2O	}	&	\cellcolor[HTML]{FBBEC1	}	2.387	&	\cellcolor[HTML]{DDEFE4	}	-0.181	&	\cellcolor[HTML]{D9E081	}	-0.164$^c$	&	\cellcolor[HTML]{FBCDCF	}	0.143$^a$	&	\cellcolor[HTML]{E9F4EF	}	-0.032$^a$	\\
\ce{S3	}	&	\cellcolor[HTML]{FBB3B6	}	2.812	&	\cellcolor[HTML]{D5ECDE	}	-0.230	&	\cellcolor[HTML]{E9E482	}	-0.113$^d$	&	\cellcolor[HTML]{FBB9BC	}	0.208$^b$	&	\cellcolor[HTML]{FBD4D7	}	0.085$^b$	\\
\ce{S4(C_{2v})	}	&	\cellcolor[HTML]{F8696B	}	5.629	&	\cellcolor[HTML]{9FD6AE	}	-0.599	&	\cellcolor[HTML]{BDD880	}	-0.252$^c$	&	\cellcolor[HTML]{F97779	}	0.429$^a$	&	\cellcolor[HTML]{FAB2B4	}	0.256$^a$	\\
\ce{CCl2	}	&	\cellcolor[HTML]{FCDFE2	}	1.117	&	\cellcolor[HTML]{F0F7F4	}	-0.053	&	\cellcolor[HTML]{FFE984	}	-0.038$^d$	&	\cellcolor[HTML]{FCF0F3	}	0.026$^b$	&	\cellcolor[HTML]{F1F7F6	}	-0.022$^b$	\\
\ce{AlCl3	}	&	\cellcolor[HTML]{FCEDF0	}	0.608	&	\cellcolor[HTML]{FCFCFF	}	0.026	&	\cellcolor[HTML]{FFDD82	}	-0.007$^c$	&	\cellcolor[HTML]{FCF6F9	}	0.005$^a$	&	\cellcolor[HTML]{FAFBFD	}	-0.006$^a$	\\
\ce{ClCN	}	&	\cellcolor[HTML]{FBD6D9	}	1.471	&	\cellcolor[HTML]{F4F8F8	}	-0.027	&	\cellcolor[HTML]{D7DF81	}	-0.170$^d$	&	\cellcolor[HTML]{FBD2D4	}	0.127$^c$	&	\cellcolor[HTML]{D6ECDE	}	-0.030$^c$	\\
\ce{OClO	}	&	\cellcolor[HTML]{FBC5C8	}	2.125	&	\cellcolor[HTML]{DEF0E5	}	-0.173	&	\cellcolor[HTML]{C3D980	}	-0.234$^c$	&	\cellcolor[HTML]{FBCDD0	}	0.143$^a$	&	\cellcolor[HTML]{ABDBB9	}	-0.087$^a$	\\
\ce{ClOO	}	&	\cellcolor[HTML]{FA9B9D	}	3.73	&	\cellcolor[HTML]{9DD5AD	}	-0.613	&	\cellcolor[HTML]{F6E883	}	-0.070$^c$	&	\cellcolor[HTML]{FA989A	}	0.318$^b$	&	\cellcolor[HTML]{FA9396	}	0.223$^b$	\\
\ce{Cl2O	}	&	\cellcolor[HTML]{FCD8DB	}	1.382	&	\cellcolor[HTML]{F1F7F5	}	-0.045	&	\cellcolor[HTML]{E3E382	}	-0.132$^d$	&	\cellcolor[HTML]{FCE0E3	}	0.080$^a$	&	\cellcolor[HTML]{CEE9D7	}	-0.029$^a$	\\
\ce{BN (^3\Pi)	}	&	\cellcolor[HTML]{FCE9EC	}	0.749	&	\cellcolor[HTML]{E0F0E7	}	-0.16	&	\cellcolor[HTML]{FFE483	}	-0.025	&	\cellcolor[HTML]{FCF1F4	}	0.024$^d$	&	\cellcolor[HTML]{FBFBFE	}	-0.001$^d$	\\
\ce{CF	}	&	\cellcolor[HTML]{FCF4F7	}	0.326	&	\cellcolor[HTML]{F1F7F6	}	-0.041	&	\cellcolor[HTML]{FFDB81	}	-0.003	&	\cellcolor[HTML]{FCFCFF	}	-0.016$^d$	&	\cellcolor[HTML]{EBF5F0	}	-0.028$^d$	\\
\ce{CH2C	}	&	\cellcolor[HTML]{FCEEF1	}	0.538	&	\cellcolor[HTML]{F4F8F8	}	-0.023	&	\cellcolor[HTML]{FFDB81	}	-0.003$^d$	&	\cellcolor[HTML]{FCF1F4	}	0.023$^c$	&	\cellcolor[HTML]{FCF4F7	}	0.024$^c$	\\
\ce{CH2CH	}	&	\cellcolor[HTML]{FCF0F3	}	0.492	&	\cellcolor[HTML]{F3F8F7	}	-0.031	&	\cellcolor[HTML]{FFDF82	}	-0.012$^d$	&	\cellcolor[HTML]{FCF2F5	}	0.020$^b$	&	\cellcolor[HTML]{FCF9FC	}	0.008$^b$	\\
\ce{C2H4	}	&	\cellcolor[HTML]{FCF0F3	}	0.491	&	\cellcolor[HTML]{F8FAFB	}	0.000	&	\cellcolor[HTML]{FFE583	}	-0.028$^d$	&	\cellcolor[HTML]{FCF0F3	}	0.026$^c$	&	\cellcolor[HTML]{FAFBFD	}	0.002$^c$	\\
\ce{CH2NH	}	&	\cellcolor[HTML]{FCEBEE	}	0.665	&	\cellcolor[HTML]{F6F9FA	}	-0.008	&	\cellcolor[HTML]{FCEA83	}	-0.052$^d$	&	\cellcolor[HTML]{FCECEF	}	0.039$^b$	&	\cellcolor[HTML]{F0F7F4	}	-0.008$^b$	\\
\ce{HCO	}	&	\cellcolor[HTML]{FCE9EC	}	0.738	&	\cellcolor[HTML]{EFF7F4	}	-0.056	&	\cellcolor[HTML]{F7E883	}	-0.068	&	\cellcolor[HTML]{FCEBEE	}	0.042$^c$	&	\cellcolor[HTML]{E4F2EB	}	-0.021$^c$	\\
\ce{H2CO	}	&	\cellcolor[HTML]{FCEBEE	}	0.653	&	\cellcolor[HTML]{F4F9F8	}	-0.022	&	\cellcolor[HTML]{F9E983	}	-0.063$^d$	&	\cellcolor[HTML]{FCEDF0	}	0.036$^c$	&	\cellcolor[HTML]{E4F2EA	}	-0.022$^c$	\\
\ce{CO2	}	&	\cellcolor[HTML]{FCDBDE	}	1.282	&	\cellcolor[HTML]{F1F7F6	}	-0.041	&	\cellcolor[HTML]{DAE081	}	-0.159$^d$	&	\cellcolor[HTML]{FCE5E8	}	0.061$^c$	&	\cellcolor[HTML]{A5D9B4	}	-0.090$^c$	\\
\ce{HNO	}	&	\cellcolor[HTML]{FCDFE2	}	1.138	&	\cellcolor[HTML]{F1F7F5	}	-0.046	&	\cellcolor[HTML]{EEE683	}	-0.096	&	\cellcolor[HTML]{FCDDE0	}	0.090$^c$	&	\cellcolor[HTML]{F6F9FA	}	0.007$^c$	\\
\ce{NO2	}	&	\cellcolor[HTML]{FBC0C3	}	2.317	&	\cellcolor[HTML]{DFF0E6	}	-0.164	&	\cellcolor[HTML]{BED880	}	-0.247$^d$	&	\cellcolor[HTML]{FBBFC2	}	0.189$^b$	&	\cellcolor[HTML]{C8E7D2	}	-0.045$^b$	\\
\ce{N2O	}	&	\cellcolor[HTML]{FBBDC0	}	2.423	&	\cellcolor[HTML]{E3F2EA	}	-0.136	&	\cellcolor[HTML]{A9D27F	}	-0.315$^d$	&	\cellcolor[HTML]{FBBDBF	}	0.197$^b$	&	\cellcolor[HTML]{94D2A5	}	-0.096$^b$	\\
\ce{O3	}	&	\cellcolor[HTML]{F98284	}	4.699	&	\cellcolor[HTML]{ABDBB9	}	-0.516	&	\cellcolor[HTML]{6CC07B	}	-0.51	&	\cellcolor[HTML]{F8696B	}	0.473$^b$	&	\cellcolor[HTML]{DBEEE2	}	0.016$^b$	\\
\ce{HOO	}	&	\cellcolor[HTML]{FCE8EA	}	0.801	&	\cellcolor[HTML]{ECF5F1	}	-0.081	&	\cellcolor[HTML]{F7E883	}	-0.066	&	\cellcolor[HTML]{FCE7EA	}	0.056$^c$	&	\cellcolor[HTML]{F3F8F7	}	0.000$^c$	\\
\ce{HOOH	}	&	\cellcolor[HTML]{FCE6E9	}	0.846	&	\cellcolor[HTML]{F6F9FA	}	-0.012	&	\cellcolor[HTML]{F1E783	}	-0.086$^d$	&	\cellcolor[HTML]{FCE6E9	}	0.059$^b$	&	\cellcolor[HTML]{E3F2EA	}	-0.018$^b$	\\
\ce{F2O	}	&	\cellcolor[HTML]{FBD2D4	}	1.638	&	\cellcolor[HTML]{EAF4EF	}	-0.092	&	\cellcolor[HTML]{DEE182	}	-0.147$^d$	&	\cellcolor[HTML]{FCDADD	}	0.098$^a$	&	\cellcolor[HTML]{D1EADA	}	-0.027$^a$	\\
\ce{HOCl	}	&	\cellcolor[HTML]{FCEBEE	}	0.682	&	\cellcolor[HTML]{F6F9FA	}	-0.011	&	\cellcolor[HTML]{FAE983	}	-0.059	&	\cellcolor[HTML]{FCEAED	}	0.045$^c$	&	\cellcolor[HTML]{EFF6F4	}	-0.005$^c$	\\
\ce{SSH	}	&	\cellcolor[HTML]{FCEBEE	}	0.684	&	\cellcolor[HTML]{F3F8F7	}	-0.03	&	\cellcolor[HTML]{FFE583	}	-0.026	&	\cellcolor[HTML]{FCECEF	}	0.038$^c$	&	\cellcolor[HTML]{FCF8FB	}	0.022$^c$	\\
\ce{B2 (^3\Sigma^-_g)	}	&	\cellcolor[HTML]{FCDCDF	}	1.254	&	\cellcolor[HTML]{D0EAD9	}	-0.27	&	\cellcolor[HTML]{F8696B	}	0.279	&	\cellcolor[HTML]{FCE3E6	}	0.069$^d$	&	\cellcolor[HTML]{F8696B	}	0.354$^c$	\\
\ce{BH	}	&	\cellcolor[HTML]{FCFBFE	}	0.047	&	\cellcolor[HTML]{F7FAFB	}	-0.002	&	\cellcolor[HTML]{FED480	}	0.014	&	\cellcolor[HTML]{FCF8FB	}	0.000$^d$	&	\cellcolor[HTML]{FCF7FA	}	0.014$^d$	\\
\ce{BH3	}	&	\cellcolor[HTML]{FCFCFF	}	0.035	&	\cellcolor[HTML]{F8FAFB	}	0.000	&	\cellcolor[HTML]{FED881	}	0.004	&	\cellcolor[HTML]{FCF8FB	}	-0.001$^d$	&	\cellcolor[HTML]{FCFBFE	}	0.004$^d$	\\
\ce{BN (^1\Sigma+)	}	&	\cellcolor[HTML]{FAA6A8	}	3.314	&	\cellcolor[HTML]{63BE7B	}	-1.005	&	\cellcolor[HTML]{B5D57F	}	-0.277	&	\cellcolor[HTML]{FAB2B4	}	0.233$^d$	&	\cellcolor[HTML]{D4ECDD	}	-0.045$^d$	\\
\ce{BF	}	&	\cellcolor[HTML]{FCF5F8	}	0.304	&	\cellcolor[HTML]{F5F9F9	}	-0.018	&	\cellcolor[HTML]{FFDD82	}	-0.008	&	\cellcolor[HTML]{FCFBFE	}	-0.010$^d$	&	\cellcolor[HTML]{ECF5F1	}	-0.017$^d$	\\
\ce{NH (^3\Sigma^-)	}	&	\cellcolor[HTML]{FCFBFE	}	0.045	&	\cellcolor[HTML]{F7FAFB	}	-0.001	&	\cellcolor[HTML]{FFD981	}	0.001	&	\cellcolor[HTML]{FCF7FA	}	0.001$^d$	&	\cellcolor[HTML]{FCFBFE	}	0.003$^d$	\\
\ce{NH2	}	&	\cellcolor[HTML]{FCF9FC	}	0.120	&	\cellcolor[HTML]{F8FAFB	}	0.001	&	\cellcolor[HTML]{FFDB81	}	-0.003	&	\cellcolor[HTML]{FCF6F9	}	0.005$^d$	&	\cellcolor[HTML]{FCFCFF	}	0.002$^d$	\\
\ce{HCN	}	&	\cellcolor[HTML]{FCE1E4	}	1.044	&	\cellcolor[HTML]{F4F8F8	}	-0.027	&	\cellcolor[HTML]{E9E482	}	-0.112	&	\cellcolor[HTML]{FCDCDF	}	0.092$^d$	&	\cellcolor[HTML]{EAF4F0	}	-0.009$^c$	\\
\ce{HOF	}	&	\cellcolor[HTML]{FCE7EA	}	0.834	&	\cellcolor[HTML]{F3F8F7	}	-0.029	&	\cellcolor[HTML]{F6E883	}	-0.072	&	\cellcolor[HTML]{FCE5E8	}	0.061$^d$	&	\cellcolor[HTML]{F2F8F6	}	-0.002$^c$	\\
\ce{AlH	}	&	\cellcolor[HTML]{FCFCFF	}	0.038	&	\cellcolor[HTML]{F7FAFB	}	-0.002	&	\cellcolor[HTML]{FED680	}	0.009	&	\cellcolor[HTML]{FCF8FB	}	0.000$^d$	&	\cellcolor[HTML]{FCF9FC	}	0.010$^d$	\\
\ce{AlH3	}	&	\cellcolor[HTML]{FCFCFF	}	0.030	&	\cellcolor[HTML]{F7FAFB	}	-0.001	&	\cellcolor[HTML]{FED781	}	0.007	&	\cellcolor[HTML]{FCF8FB	}	0.000$^d$	&	\cellcolor[HTML]{FCFAFD	}	0.007$^d$	\\
\ce{AlF	}	&	\cellcolor[HTML]{FCF6F9	}	0.241	&	\cellcolor[HTML]{F6F9FA	}	-0.011	&	\cellcolor[HTML]{FFDD82	}	-0.008	&	\cellcolor[HTML]{FCFBFE	}	-0.012$^d$	&	\cellcolor[HTML]{EAF4EF	}	-0.021$^d$	\\
\ce{AlCl	}	&	\cellcolor[HTML]{FCF4F7	}	0.310	&	\cellcolor[HTML]{F7FAFB	}	-0.001	&	\cellcolor[HTML]{FFD981	}	0.001	&	\cellcolor[HTML]{FCF6F8	}	0.007$^d$	&	\cellcolor[HTML]{FCF9FC	}	0.009$^d$	\\
\ce{SiH	}	&	\cellcolor[HTML]{FCFCFF	}	0.019	&	\cellcolor[HTML]{F7FAFB	}	-0.001	&	\cellcolor[HTML]{FFDA81	}	-0.001	&	\cellcolor[HTML]{FCF8FB	}	0.000$^d$	&	\cellcolor[HTML]{FBFBFE	}	0.000$^d$	\\
\ce{SiH4	}	&	\cellcolor[HTML]{FCFCFF	}	0.032	&	\cellcolor[HTML]{F8FAFB	}	0.001	&	\cellcolor[HTML]{FFDB81	}	-0.002	&	\cellcolor[HTML]{FCF8FB	}	-0.001$^d$	&	\cellcolor[HTML]{F8FAFC	}	-0.003$^d$	\\
\hline\hline
\end{tabular}
}
\end{table}

\addtocounter{table}{-1}
\begin{table}
\caption{(Continued)}

{\footnotesize
\begin{tabular}{clllll}
\hline\hline
\ce{SiO	}	&	\cellcolor[HTML]{FCE1E4	}	1.042	&	\cellcolor[HTML]{E7F3ED	}	-0.113	&	\cellcolor[HTML]{F3E783	}	-0.080	&	\cellcolor[HTML]{FCECEF	}	0.038$^d$	&	\cellcolor[HTML]{D6ECDE	}	-0.042$^d$	\\
\ce{SiF	}	&	\cellcolor[HTML]{FCF6F9	}	0.238	&	\cellcolor[HTML]{F5F9F9	}	-0.017	&	\cellcolor[HTML]{FFE182	}	-0.017	&	\cellcolor[HTML]{FCFCFF	}	-0.013$^d$	&	\cellcolor[HTML]{E1F1E8	}	-0.031$^d$	\\
\ce{CS	}	&	\cellcolor[HTML]{FCDFE2	}	1.133	&	\cellcolor[HTML]{EFF6F4	}	-0.059	&	\cellcolor[HTML]{FBE983	}	-0.056	&	\cellcolor[HTML]{FCE1E3	}	0.077$^d$	&	\cellcolor[HTML]{FCF4F7	}	0.027$^d$	\\
\ce{H2	}	&	\cellcolor[HTML]{FCFCFF	}	0.000	&	\cellcolor[HTML]{F8FAFB	}	0.000	&	\cellcolor[HTML]{FFDA81	}	0.000	&	\cellcolor[HTML]{FCF8FB	}	0.000$^d$	&	\cellcolor[HTML]{FCFCFF	}	0.000$^d$	\\
\ce{OH	}	&	\cellcolor[HTML]{FCFAFD	}	0.093	&	\cellcolor[HTML]{F8FAFB	}	0.000	&	\cellcolor[HTML]{FFDC82	}	-0.005	&	\cellcolor[HTML]{FCF7FA	}	0.003$^d$	&	\cellcolor[HTML]{FAFBFD	}	-0.002$^d$	\\
\ce{HF	}	&	\cellcolor[HTML]{FCF9FC	}	0.126	&	\cellcolor[HTML]{F8FAFB	}	0.002	&	\cellcolor[HTML]{FFE082	}	-0.016	&	\cellcolor[HTML]{FCF7F9	}	0.004$^d$	&	\cellcolor[HTML]{F1F7F6	}	-0.012$^d$	\\
\ce{H2O	}	&	\cellcolor[HTML]{FCF7FA	}	0.225	&	\cellcolor[HTML]{F8FAFC	}	0.003	&	\cellcolor[HTML]{FFE483	}	-0.026	&	\cellcolor[HTML]{FCF4F7	}	0.011$^d$	&	\cellcolor[HTML]{EFF6F3	}	-0.014$^d$	\\
\ce{CH (^3\Pi)	}	&	\cellcolor[HTML]{FCFCFF	}	0.033	&	\cellcolor[HTML]{F8FAFB	}	0	&	\cellcolor[HTML]{FED881	}	0.004	&	\cellcolor[HTML]{FCF8FB	}	0.001$^d$	&	\cellcolor[HTML]{FCFAFD	}	0.005$^d$	\\
\ce{CH2 (^3B_1)	}	&	\cellcolor[HTML]{FCFCFF	}	0.033	&	\cellcolor[HTML]{F8FAFB	}	0.001	&	\cellcolor[HTML]{FFDA81	}	-0.001	&	\cellcolor[HTML]{FCF7FA	}	0.001$^d$	&	\cellcolor[HTML]{FBFBFE	}	0.000$^d$	\\
\ce{CH3	}	&	\cellcolor[HTML]{FCFBFE	}	0.064	&	\cellcolor[HTML]{F8FAFB	}	0.002	&	\cellcolor[HTML]{FFDB81	}	-0.002	&	\cellcolor[HTML]{FCF7FA	}	0.002$^d$	&	\cellcolor[HTML]{FBFBFE	}	0.000$^d$	\\
\ce{CH4	}	&	\cellcolor[HTML]{FCFAFD	}	0.095	&	\cellcolor[HTML]{F8FAFC	}	0.003	&	\cellcolor[HTML]{FFDB81	}	-0.003	&	\cellcolor[HTML]{FCF7FA	}	0.001$^d$	&	\cellcolor[HTML]{F9FBFD	}	-0.002$^d$	\\
\ce{CCH	}	&	\cellcolor[HTML]{FCE8EA	}	0.804	&	\cellcolor[HTML]{EFF6F4	}	-0.059	&	\cellcolor[HTML]{FFDA81	}	-0.001	&	\cellcolor[HTML]{FCE5E8	}	0.063$^c$	&	\cellcolor[HTML]{FCE2E5	}	0.066$^c$	\\
\ce{C2H2	}	&	\cellcolor[HTML]{FCE8EB	}	0.796	&	\cellcolor[HTML]{F6F9F9	}	-0.013	&	\cellcolor[HTML]{FAE983	}	-0.058	&	\cellcolor[HTML]{FCE4E7	}	0.065$^c$	&	\cellcolor[HTML]{FCFAFD	}	0.014$^c$	\\
\ce{NH3	}	&	\cellcolor[HTML]{FCF7FA	}	0.200	&	\cellcolor[HTML]{F8FAFC	}	0.004	&	\cellcolor[HTML]{FFE082	}	-0.015	&	\cellcolor[HTML]{FCF5F8	}	0.009$^d$	&	\cellcolor[HTML]{F7FAFA	}	-0.005$^d$	\\
\ce{C2	}	&	\cellcolor[HTML]{FAA3A5	}	3.441	&	\cellcolor[HTML]{A0D6AF	}	-0.592	&	\cellcolor[HTML]{63BE7B	}	-0.539	&	\cellcolor[HTML]{F98A8C	}	0.365$^d$	&	\cellcolor[HTML]{63BE7B	}	-0.171$^d$	\\
\ce{N2	}	&	\cellcolor[HTML]{FCDEE0	}	1.184	&	\cellcolor[HTML]{F4F8F8	}	-0.026	&	\cellcolor[HTML]{E1E282	}	-0.137	&	\cellcolor[HTML]{FBD5D8	}	0.115$^d$	&	\cellcolor[HTML]{E9F4EE	}	-0.018$^d$	\\
\ce{CO	}	&	\cellcolor[HTML]{FCEAED	}	0.722	&	\cellcolor[HTML]{F4F8F8	}	-0.027	&	\cellcolor[HTML]{F6E883	}	-0.069	&	\cellcolor[HTML]{FCEBEE	}	0.044$^d$	&	\cellcolor[HTML]{E5F2EB	}	-0.024$^d$	\\
\ce{CN	}	&	\cellcolor[HTML]{FBD4D6	}	1.566	&	\cellcolor[HTML]{C9E7D3	}	-0.314	&	\cellcolor[HTML]{DBE081	}	-0.155	&	\cellcolor[HTML]{FBD3D5	}	0.123$^d$	&	\cellcolor[HTML]{DFF0E6	}	-0.028$^d$	\\
\ce{NO	}	&	\cellcolor[HTML]{FCE2E5	}	1.011	&	\cellcolor[HTML]{EEF6F3	}	-0.066	&	\cellcolor[HTML]{ECE582	}	-0.102	&	\cellcolor[HTML]{FCDCDF	}	0.092$^d$	&	\cellcolor[HTML]{F3F8F7	}	-0.007$^d$	\\
\ce{O2	}	&	\cellcolor[HTML]{FCDEE0	}	1.183	&	\cellcolor[HTML]{F3F8F7	}	-0.032	&	\cellcolor[HTML]{E9E482	}	-0.112	&	\cellcolor[HTML]{FBD5D7	}	0.117$^d$	&	\cellcolor[HTML]{FCFAFD	}	0.009$^d$	\\
\ce{OF	}	&	\cellcolor[HTML]{FCE9EC	}	0.734	&	\cellcolor[HTML]{E2F1E9	}	-0.144	&	\cellcolor[HTML]{FCEA83	}	-0.051	&	\cellcolor[HTML]{FCEBEE	}	0.041$^d$	&	\cellcolor[HTML]{F3F8F7	}	-0.008$^d$	\\
\ce{F2	}	&	\cellcolor[HTML]{FCE2E5	}	1.019	&	\cellcolor[HTML]{F0F7F4	}	-0.052	&	\cellcolor[HTML]{F4E883	}	-0.076	&	\cellcolor[HTML]{FCE1E3	}	0.077$^d$	&	\cellcolor[HTML]{FCFCFF	}	0.003$^d$	\\
\ce{PH3	}	&	\cellcolor[HTML]{FCFAFD	}	0.108	&	\cellcolor[HTML]{F7FAFB	}	-0.002	&	\cellcolor[HTML]{FFDA81	}	-0.002	&	\cellcolor[HTML]{FCF7FA	}	0.001$^d$	&	\cellcolor[HTML]{FBFBFE	}	0.002$^d$	\\
\ce{HS	}	&	\cellcolor[HTML]{FCFBFE	}	0.067	&	\cellcolor[HTML]{F8FAFB	}	0.001	&	\cellcolor[HTML]{FFDA81	}	0.000	&	\cellcolor[HTML]{FCF7FA	}	0.002$^d$	&	\cellcolor[HTML]{FCFCFF	}	0.003$^d$	\\
\ce{H2S	}	&	\cellcolor[HTML]{FCF8FB	}	0.16	&	\cellcolor[HTML]{F8FAFB	}	0.002	&	\cellcolor[HTML]{FFDA81	}	-0.001	&	\cellcolor[HTML]{FCF6F9	}	0.005$^d$	&	\cellcolor[HTML]{FCFBFE	}	0.006$^d$	\\
\ce{HCl	}	&	\cellcolor[HTML]{FCFAFD	}	0.111	&	\cellcolor[HTML]{F8FAFC	}	0.003	&	\cellcolor[HTML]{FFDB81	}	-0.003	&	\cellcolor[HTML]{FCF7FA	}	0.004$^d$	&	\cellcolor[HTML]{FCFCFF	}	0.002$^d$	\\
\ce{SO	}	&	\cellcolor[HTML]{FCE3E6	}	0.965	&	\cellcolor[HTML]{F1F7F6	}	-0.043	&	\cellcolor[HTML]{EDE582	}	-0.101	&	\cellcolor[HTML]{FCE0E3	}	0.079$^d$	&	\cellcolor[HTML]{E9F4EE	}	-0.018$^d$	\\
\ce{ClO	}	&	\cellcolor[HTML]{FCE9EB	}	0.763	&	\cellcolor[HTML]{EBF5F0	}	-0.088	&	\cellcolor[HTML]{FFEB84	}	-0.043	&	\cellcolor[HTML]{FCE7EA	}	0.055$^d$	&	\cellcolor[HTML]{FCF7FA	}	0.015$^d$	\\
\ce{ClF	}	&	\cellcolor[HTML]{FCEFF1	}	0.536	&	\cellcolor[HTML]{F4F9F8	}	-0.022	&	\cellcolor[HTML]{FFE784	}	-0.032	&	\cellcolor[HTML]{FCEFF2	}	0.028$^d$	&	\cellcolor[HTML]{F8FAFB	}	-0.003$^d$	\\
\ce{P2	}	&	\cellcolor[HTML]{FBD2D5	}	1.616	&	\cellcolor[HTML]{ECF5F1	}	-0.081	&	\cellcolor[HTML]{F3E783	}	-0.081	&	\cellcolor[HTML]{FBC7CA	}	0.161$^d$	&	\cellcolor[HTML]{FCDBDE	}	0.089$^d$	\\
\ce{S2	}	&	\cellcolor[HTML]{FCE4E7	}	0.948	&	\cellcolor[HTML]{F3F8F7	}	-0.030	&	\cellcolor[HTML]{FEEA83	}	-0.045	&	\cellcolor[HTML]{FCE0E2	}	0.080$^d$	&	\cellcolor[HTML]{FCEEF1	}	0.042$^d$	\\
\ce{Cl2	}	&	\cellcolor[HTML]{FCEFF2	}	0.520	&	\cellcolor[HTML]{F7FAFB	}	-0.003	&	\cellcolor[HTML]{FFE383	}	-0.022	&	\cellcolor[HTML]{FCEEF1	}	0.034$^d$	&	\cellcolor[HTML]{FCF7FA	}	0.015$^d$	\\
\hline\hline
\end{tabular}
 
}
(a) cc-pVDZ/ VDZ(d,p)

(b) cc-pVTZnoF/ VTZ(d,p)

(c) cc-pVTZ/VTZ(f,d) 
 
(d) cc-pVQZnoG/VQZ(f,d)

(e) remaining values in T4-(Q)$_\Lambda$ column are at cc-pVQZ level 
\end{table}

A small, but not decisive, advantage is seen for the Neese-Valeev ano-pVDZ over standard cc-pVDZ, but ano-pVTZ does seem to recover substantially more of the (Q) than standard cc-pVTZ.

\begin{table*}[!htbp]
\centering
\caption{Statistics of TAE$_e$ contributions (\kcalmol) for the W4-08 dataset. \label{tab:W4-08}}
\renewcommand{\arraystretch}{0.9}
\begin{tabular}{lccccccccc}
\hline\hline
\multicolumn{1}{l}{\textbf{RMSD from CBS limit}}                & {\textbf{ano-pVDZ}} & {\textbf{ano-pVTZ}} &{\textbf{cc-pVDZ(p,s)}}& {\textbf{cc-pVDZ}} & {\textbf{cc-pVTZ(d,p)}} & {\textbf{cc-pVTZ}} & {\textbf{cc-pVQZ(f,d)}} & {\textbf{cc-pVQZ}}   & {\textbf{cc-pV\{T,Q\}Z}} \\
\hline
\textbf{CCSDT(Q)--CCSDT}                                                                   & {0.282}    & {0.058}    && {0.331}    & {0.233}     & {0.101}    & {0.027}    & {REF}   & {0.061}     \\
\textbf{relative to \{T,Q\}}                                                   &  {0.335}   & {0.119}   && {0.387}    & {0.293}     & {0.162}   & {0.087}   & {0.061} & {REF}       \\
\textbf{CCSDTQ--CCSDT(Q)}                                                                & {0.024}    & {0.009}    &{0.099}& {0.034}    & {0.019}     & {0.015}  & {0.004}    & {REF} & {0.013}    \\
\textbf{relative to \{T,Q\}}                                                   & {0.028}   & {0.022}    &{0.108}& {0.046}   & {0.030}     & {0.028}   & {0.016}    & {0.013} & {REF}      \\
\textbf{CCSDT(Q)$_\Lambda$--CCSDT(Q)}                                       & {0.023}    & {0.006}   &{0.080}& {0.034}   & {0.015}    & {0.009}    & {0.002}    & {REF}   & {\textbf{}}          \\
\textbf{without BN}                                                            & {0.022}   & {0.005}    &{0.060}& {0.030}    & {0.015}     & {0.008}    & {0.002}    & {REF}   & {\textbf{}}          \\
\textbf{CCSDTQ--CCSDT(Q)$_\Lambda$}                                      & {0.025}   & {0.007}    &{0.063}& {0.018}    & {0.012}     & {0.010}    & {0.003}    &{REF}   & {\textbf{}}          \\
\textbf{CCSDTQ(5)$_\Lambda$ -- CCSDTQ}                                           & \textbf{}                             & \textbf{}                             &{0.023}& {0.011}    & {0.007}    & {REF}     & {0.005}                             & \textbf{}                          & \textbf{}                              \\
\textbf{without BN}                                                            & \textbf{}                             & \textbf{}                             &{0.023}& {0.010}    & {0.006}     & {REF}      & {0.004}                            & \textbf{}                          & \textbf{}                              \\
\textbf{CCSDTQ(5)$_\Lambda$ -- CCSDT(Q)$_\Lambda$}             & \textbf{}                             & \textbf{}                             &{0.063}& {0.018}    & {0.014}     & {REF}      & {0.003}                            & \textbf{}                          & \textbf{}                              \\
\textbf{without BN}                                                            & {\textbf{}}         & {\textbf{}}         &{0.062}& {0.014}    & {0.012}    & {REF}      & {0.002}         & {\textbf{}}      & {\textbf{}}          \\
\textbf{}                                                                      & {\textbf{}}         & {\textbf{}}         & {\textbf{}}         & {\textbf{}}          & {\textbf{}}         & {\textbf{}}         & {\textbf{}}      & {\textbf{}}          \\
                                                                               &                                       & {\textbf{}}         & {\textbf{}}         & {\textbf{}}          & {\textbf{}}         & {\textbf{}}         & {\textbf{}}      & {\textbf{}}          \\
\hline\multicolumn{8}{c}{\textbf{MEAN SIGNED VALUE}}   \\\hline

\textbf{CCSDT(Q)--CCSDT}                                                                    & 0.855                                 & 0.947                                 && 0.817                                 & 0.833                                  & 0.917                                 & 0.972                                 & 0.989                              & 1.033                                  \\
\textbf{CCSDTQ--CCSDT(Q)}                                                                & -0.181                                & -0.152                                &-0.116& -0.149                                & -0.158                                 & -0.149                                & -0.136                                & -0.125                             & -0.131                                 \\\textbf{CCSDT(Q)$_\Lambda$--CCSDT(Q)}                                        & -0.080                                & -0.084                                &-0.054& -0.075                                & -0.080                                 & -0.082                                & -0.083                                & -0.080                             &                                        \\
\textbf{CCSDTQ--CCSDT(Q)$_\Lambda$}                                      & -0.101                                & -0.072                                &-0.062& -0.074                                & -0.078                                 & -0.070                                & -0.066                                & -0.052                             &                                        \\
\textbf{CCSDTQ(5)$_\Lambda$ -- CCSDTQ}                                           &                                       &                                       &0.055& 0.066                                 & 0.058                                  & 0.046                                 &   0.041                                    &                                    & \multicolumn{1}{c}{}                   \\
\textbf{CCSDTQ(5)$_\Lambda$ -- CCSDT(Q)$_\Lambda$}             &                                       &                                       &-0.007& -0.008                                & -0.005                                 & 0.003                                & -0.005                                      &                                    &                                        \\
                                                                               &                                       &                                       &                                       &                                        &                                       &                                       &                                    &                                        \\
\hline\multicolumn{8}{c}{\textbf{RMS VALUE for W4-08 or available points}}   \\\hline
\textbf{CCSDT(Q)--CCSDT}                                                                   & 1.248                                 & 1.404                                 && 1.188                                 & 1.247                                  & 1.362                                 & 1.436                                 & 1.459                              & 1.518                                  \\
\textbf{CCSDTQ--CCSDT(Q)}                                                               & 0.308                                 & 0.284                                 &0.247& 0.269                                 & 0.290                                  & 0.278                                 & 0.268                                 & 0.287                             & 0.297                                  \\
\textbf{CCSDT(Q)$_\Lambda$--CCSDT(Q)}                                        & 0.169                                 & 0.183                                 &0.118& 0.157                                 & 0.176                                  & 0.180                                 & 0.185                                 & 0.182                              &                                        \\
\textbf{CCSDTQ--CCSDT(Q)$_\Lambda$}                                      & 0.161                                 & 0.132                                 &0.149& 0.132                                 & 0.138                                  & 0.128                                 & 0.126                                 & 0.124                              &                                        \\
\textbf{CCSDTQ(5)$_\Lambda$ -- CCSDTQ}                                           &                                       &                                       &0.097& 0.113                                 & 0.100                                  & 0.076                                 &     0.082                                  &                                    & \multicolumn{1}{c}{}                   \\
\textbf{CCSDTQ(5)$_\Lambda$ -- CCSDT(Q)$_\Lambda$}             &                                       &                                       &0.096& 0.067                                 & 0.064                                  & 0.056                                 & 0.062                                      &                                    &                                       
\\\hline\hline
\end{tabular}

Karton's extrapolation formula\cite{Karton2020} was employed for the \{T,Q\} extrapolations.

\end{table*}
The $(Q)_\Lambda - (Q)$ difference appears to converge fairly rapidly and smoothly with the basis set, mean signed averages hovering around -0.08 \kcalmol\ across basis sets. 
Even the smallish VTZ(d,p) basis set can achieve 0.015 \kcalmol\ RMS from the cc-pVQZ result. 

As for $T_4 - (Q)_\Lambda$, even cc-pVDZ can come within 0.014 \kcalmol\ RMS. Like $(Q)_\Lambda - (Q)$, $T_4 - (Q)_\Lambda$ slightly reduces atomization energies; based on the mean signed values, smaller basis sets tend to exaggerate the effect in absolute value.

We only have limited quintuples data, confined to the cc-pVDZ(p,s)cc-pVDZ, cc-pVTZ(d,p), and (for a subset) cc-pVTZ basis sets, plus cc-pVQZ(f,d) for diatomics. (The latter, however, do include such troublesome species as \ce{B2}, \ce{C2}, and \ce{BN}, plus \ce{P2} with its notoriously\cite{PerssonP2P4,jmlm207} slow basis set convergence.) As already found previously,\cite{jmlm330} $(5)_\Lambda$ systematically increases TAE, and the magnitude of the effect is comparable to the \emph{decrease} in TAE from $T_4 - (Q)_\Lambda$. Moreover, the basis set expansion effects appear to run in counterphase, which invites the question how $Q(5)_\Lambda - (Q)_\Lambda$ would behave. (With the cc-pVDZ basis set, this is the correction term that upgrades `W5preview1' to `W5preview2' in our recent paper.\cite{jmlm340}) 

Indeed, as we see in Table~\ref{tab:W4-08}, the mean signed contribution is nearly zero, while its RMS value is a modest 0.06 \kcalmol.
For cc-pVDZ, the RMS difference from cc-pVTZ is 0.018 \kcalmol, which drops to 0.014 \kcalmol\ upon omission of \ce{BN}. The RMS difference between cc-pVTZ and cc-pVQZ(f,d) is just 0.002 \kcalmol, illustrating that cc-pVTZ is practically at the basis set limit for CCSDTQ(5)$_\Lambda$--CCSDT(Q)$_\Lambda$.

Figure 1 has a box-and-whiskers plot of the various contributions: one sees there that $Q(5)_\Lambda - (Q)_\Lambda$ does have a quite narrow distribution. A few positive outliers are \ce{B2}, \ce{S4}, and \ce{ClOO}, while a few negative outliers are \ce{C2}, \ce{BN}, and \ce{SO3}.

For a subset of systems, we also have fully iterative CCSDTQ5/cc-pVDZ data. Their RMSD with CCSDTQ(5)$_\Lambda$/cc-pVDZ is just 0.01 kcal/mol.

If one were aiming for 0.01 \kcalmol\ accuracy for post-CCSDT(Q) correlation effects, one could assemble a composite of [CCSDT(Q)$_\Lambda$--CCSDT(Q)]/cc-pVTZ with [CCSDTQ(5)$_\Lambda$--CCSDT(Q)$_\Lambda$]/cc-pVTZ(d,p), while a more reasonable 0.1 \kJmol (0.024 \kcalmol) could be achieved with [CCSDT(Q)$_\Lambda$--CCSDT(Q)]/cc-pVTZ(d,p) or even ano-pVDZ combined with [CCSDTQ(5)$_\Lambda$--CCSDT(Q)$_\Lambda$]/cc-pVDZ.

One of our working assumptions when we set out on this investigation was that (based on Ref.\citenum{jmlm205}) basis set convergence of the (very costly) connected quintuples would be considerably faster than for the higher-order connected quadruples --- and that we could exploit this in a composite scheme in which CCSDTQ--CCSDT(Q) or CCSDTQ--CCSDT(Q)$_\Lambda$ would be treated in a larger basis set than (5)$_\Lambda$. The present study, at first sight, appears to militate in favor of a `one-shot' CCSDTQ(5)$_\Lambda$--CCSDT(Q)$_\Lambda$ correction. 

The mind wonders whether post-CCSDT(Q) corrections with the unpolarized cc-pVDZ(p,s) basis set would be of any practical use, given that they can be obtained for far larger systems than those with the full cc-pVDZ basis set. As it happens, performance for the especially costly (5)$_\Lambda$ contribution is surprisingly decent, RMSD=0.023 \kcalmol, compared with an RMS contribution of 0.097 \kcalmol\ for that basis set. However, performance with that unpolarized basis set for (Q)$_\Lambda$ -(Q) is quite disappointing, RMSD=0.080 \kcalmol\ out of an RMS contribution of 0.118 \kcalmol. For CCSDTQ-CCSDT(Q)$_\Lambda$, things are not much better, 0.061 out of 0.15 \kcalmol. 

This appears to be one scenario where a composite post-CCSDT(Q)$_\Lambda$ correction is beneficial --- specifically, combining CCSDTQ/cc-pVDZ using the rapid CFOUR code with CCSDTQ(5)$_\Lambda$/cc-pVDZ(p,s) using the general (but slower) MRCC code. The RMSD we find for this composite is a tolerable 0.030 \kcalmol\ --- although it should be kept in mind that the RMS \emph{contribution} is just 0.06 \kcalmol\ with a signed average of almost exactly zero -- if the remaining error sources in other calculation steps of one's thermochemical protocol have cumulative expected errors larger than 0.06, then perhaps one should stop at CCSDT(Q)$_\Lambda$ and ``call it a day''.

Combining CCSDTQ/cc-pVTZ(d,p) and CCSDTQ(5)$_\Lambda$/cc-pVDZ basis sets in the same manner is comparable in accuracy to just the single-shot cc-pVDZ correction. CCSDTQ/cc-pVTZ with CCSDTQ(5)$_\Lambda$/cc-pVDZ is somewhat more accurate, at RMSD=0.009 \kcalmol\ excluding BN --- however, one might wish to consider whether one can truly tighten all other potential error sources to better than 0.01 \kcalmol\ before making this computer time investment. A truncated cc-pVTZ basis set for quintuples combined with the full cc-pVTZ basis set for full quadruples reduces RMSD further to 0.006 \kcalmol\ --- where the same question about computational cost-benefit ratio becomes even more acute.

Very recently, we demonstrated\cite{jmlm344} another viable alternative for the quintuples: FNO-CCSDTQ$(5)_\Lambda$ (frozen natural orbital coupled cluster\cite{Taube2005} theory as implemented\cite{RolikKallay2011} in MRCC\cite{MRCC2025,MRCCcode}), for such high-order terms, appears to converge very rapidly with the FNO truncation cutoff, and useful answers for $(5)_\Lambda$ can be obtained with cutoffs as coarse as 0.001.


\subsection{A note on dealing with UHF bifurcation}

\jmlm{For two of the radical systems in the W4-08 dataset, namely, \ce{FOO} and \ce{ClOO}, bifurcation of the UHF solution has previously been documented by Denis and Ornellas.\cite{Denis2008SpinContaminationXOO} (This is actually a special case of the whimsically named `triplet instability in doublets' described by Stanton and coworkers.\cite{Szalay2004TripletInstabilityDoublets}) In both cases, one solution (`LowS2') has an $<S^2>$ near the pure doublet value of 0.75, while the other (`HighS2') has a much lower SCF energy and an elevated $<S^2>$ of 1.52 for \ce{FOO} and 1.55 for \ce{ClOO}. 
}

\jmlm{As discussed in detail in the Supplementary Material and shown in Table S1 there, energies with `LowS2' and `ROHF' references are much closer to each other than either is to `HighS2', while the latter exhibits anomalously large maximum $T_1$ amplitudes and otherwise behaves erratically. We were able to reliably converge to the `LowS2' UHF solution by the expedient of reading in the density from a converged ROHF calculation as the initial guess. }

\subsection{Timing data for open-shell CCSDTQ in CFOUR}

How well does this code parallelize? Table~\ref{tab:timings} presents some CCSDTQ/cc-pVTZ timing data as a function of the number of cores for the \ce{NO2} radical and the \ce{N2O} linear triatomic, all run on identical and otherwise empty Intel Ice Lake nodes.

The open-shell case parallelizes tolerably well through 8 cores, and a speedup is still seen for 16 cores, but wall clock times for 24 and 32 cores actually \emph{increase}, strongly suggesting that memory bus contention has set in.

In contrast, for the closed-shell case, parallelization is almost perfect through 8 cores and tolerably good through 16 cores, while even 24 cores still shows a marked further speedup. Only at 32 cores does contention appear to have set in.

That said, for systems with more electrons, such as \ce{P4}, parallelization over 32 cores still appears to be fairly efficient. We do note that run times for this system with 4/3 as many basis functions and 5/4 as many valence electrons are an order of magnitude longer.

How does CFOUR compare with another CCSDTQ code? By way of illustration, for \ce{NO2}, the first CCSDTQ/cc-pVTZ(d,p) iteration on 16 cores takes 
4110 seconds wall-clock with MRCC, and just 923 seconds with CFOUR. The performance gap is much greater for closed-shell \ce{N2O}: 4367 seconds wall time with MRCC, and just 236 seconds with CFOUR on the same number of cores. This is not to detract from the achievements that MRCC represents, but to illustrate the performance differences between a completely general coupled-cluster code for arbitrary excitation levels versus one that was hand-coded for a specific excitation level.

\begin{table}

\caption{Wall clock times (s) with various numbers of CPU cores for the first CCSDTQ/cc-pVTZ iteration on an open-shell case and two closed-shell cases\label{tab:timings}}
\begin{tabular}{rrrrr}
\hline\hline
$N_\mathrm{cores}$ & \ce{NO2} & \ce{N2O} & \ce{P4} & \ce{S4}\\
\hline
\multicolumn{5}{c}{Default core binding}\\
\hline
1 & 16207 & 8127 & 85363&128644\\
2 & 9845 & 4134 & 43078 & 65798\\
4 & 5485 & 2119 & 22653 & 34044\\
8  & 3873 & 1124 & 11327 & 18542\\
16  & 2642 & 653 & 7661&11820\\ 
24  & 3072 & 523 & 5020 & 8367\\
32 & 3359 & 754 & 4627 & 7063\\
48 &  ---     & ---    & 6917 & 7387\\
\hline

\multicolumn{5}{c}{Cores packed onto 1 socket}\\
\hline
8 & 3630 & 1138&12589&\\
16 & 2648 & 669 & 6365\\
24 & 2435 & 547 & 5027 \\
\hline
\multicolumn{5}{c}{Cores spread over 2 sockets}\\
\hline
8 & 3967 & 1378&12259\\
16 & 3084 & 767 & \\
24 & 2953 & 633 &  5475 \\
\hline\hline      
\end{tabular}

Jobs were run on dedicated dual-socket Intel(R) Xeon(R) Gold 5320 CPUs at 2.20GHz with 256GB or more RAM, except \ce{S4} which did not fit in available RAM and needed to be run on an AMD Zen5 machine with 1.5TB RAM.


\end{table}

As NCC with TBLIS is quite memory bandwidth-sensitive, in a NUMA (non-uniform memory access) system like our compute nodes, potentially a slight speedup might be realized by forcibly `packing' all processes onto a single socket. Indeed, timing comparison between such `packing' and forcibly `spreading' over both sockets reveals that the latter comes at a 10-20\% performance penalty (Table~\ref{tab:timings}). (For low core counts, however, this comparison will be skewed since a die with only a few active cores may go to a higher clock frequency --- TurboBoost cannot be disabled in user space on any HPC system we have access to.)

\subsection{Proof of principle: electron affinity of ozone}

The strong static correlation in ozone is quite well known(e.g., Ref.\citenum{jmlm326} and references therein), resulting from a near-degeneracy between HOMO and LUMO that causes a double HOMO-LUMO excitation to have a $T_2$ amplitude of almost 0.3. In contrast, static correlation is less strong, and of a different nature, in the anion: at the CCSDTQ/jun-cc-pVDZ level, we found a largest $T_1$ amplitude of almost 0.14 but no similarly prominent double excitations. As such, the electron affinity will be strongly dependent on the level of theory, as no error compensation can be expected. 

Neumark and coworkers\cite{Arnold1994} measured an adiabatic EA of 2.103$\pm$0.003 eV, affirming earlier work by Lineberger and coworkers\cite{Novick1979}. A classic 1994 paper by Roos and coworkers\cite{Roos1993OzoneEA} applied their then-newly developed CASPT2 multireference method\cite{Andersson1990CASPT2,Andersson1992CASPT2} and obtained 2.19 eV, including a zero-point correction of 0.023 eV which we shall reuse here. Musia{\l} et al. obtained\cite{Musial2009} 1.64 eV at the EOM-EA-CCSDT/aug-cc-pVTZ level\cite{nooijen1995eom-ea} (for EOM-coupled cluster and EOM-CCSDT see Refs.\citenum{stanton1993eomcc} and 
\citenum{kucharski1993eommethods}, 
respectively). (This is, however, a vertical EA which cannot be directly compared, other than as an approximate lower limit to the adiabatic EA.
At the CCSDT(Q)$_\Lambda$/jun-cc-pVTZ level, we presently calculate a vertical-adiabatic EA difference of 0.39 eV.)

In the present paper, we applied a variant of the W5preview2 protocol\cite{jmlm340}. The UCCSD(T)/aug-cc-pCV\{5,6\}Z extrapolated basis set limit is 2.1395 eV; our best estimate of the CCSDT(Q)$_\Lambda$-CCSD(T) difference was obtained with the aug-cc-pVQZ basis set, and is -0.0525 eV. With great difficulty and allocating over 1.4 TB of RAM to the anion calculation, we were able to bring CCSDTQ/jun-cc-pVTZ to completion: the resulting best estimate for the CCSDTQ--CCSDT(Q)$_\Lambda$ correction is a surprisingly modest 0.0088 eV, almost identical to the 0.0091 eV obtained with the smaller aug-cc-pVDZ basis set.

For the asymmetric quintuples correction $(5)_\Lambda$ we availed ourselves of the MRCC\cite{MRCC2025,MRCCcode} program system: with the jun-cc-pVDZ basis set, we obtained -0.0109 eV, once again highlighting the degree of error compensation from which the CCSDT(Q)$_\Lambda$ method benefits\cite{jmlm326,jmlm330}. A scalar relativistic correction of -0.0060 eV was obtained using the X2C (exact two-component\cite{ReiherX2C}) approach with the ACVQZ basis set, which result agrees to 4 decimal places with the smaller basis set ACVTZ.  

At the end of the day, applying the zero-point energy from Roos and coworkers, we find EA(\ce{O3})=2.102 eV, in pleasing agreement with experiment. In a future study, we shall revisit the potential energy surfaces of both species and hence update the ZPVE, as we expect to be able to achieve a smaller uncertainty than experiment.

\section{Conclusions}\label{sec:conclusions}
In the present note, we report the extension of the CCSDTQ implementation in CFOUR to UHF and ROHF references, and demonstrate its computational efficiency.

As an application, we study the basis set convergence of post-CCSDT(Q) corrections for the W4-08 thermochemical dataset in detail. Basis set convergence for (Q)$_\Lambda$--(Q) is comparatively fast. While there are indications that for tricky species such as \ce{B2} or \ce{O3}, basis set convergence of CCSDTQ--CCSDT(Q)$_\Lambda$ is slower than that of (5)$_\Lambda$, both the effects and their basis set dependences work against each other. As a result, a single-shot CCSDTQ(5)$_\Lambda$-CCSDT(Q)$_\Lambda$ correction appears to be the most efficient way to proceed. At least a cc-pVDZ basis set is required for that; the unpolarized cc-pVDZ(p,s) basis set is inadequate. That said, a combination of CCSDTQ/cc-pVDZ and CCSDTQ(5)$_\Lambda$/cc-pVDZ(p,s) might be viable as a reduced-cost alternative. \jmlm{A better alternative than such aggressive basis set truncation, as shown in Ref.\cite{jmlm344}, might be to substitute $(5)_\Lambda$/cc-pVDZ evaluated in a small subset of frozen natural orbitals.}

 As a by-product of the present study: at least for the examples of \ce{FOO} and \ce{ClOO} encountered here, if the UHF solution bifurcates into `LowS2' and `HighS2', energetics for the `LowS2' solution are clearly more stable, as well as closer to the ROHF solutions.

Finally, our best computed adiabatic electron affinity of ozone is in excellent agreement with experiment.

\subsection*{CRediT authorship contribution statement}
GHJ: programming (lead), conceptualization (supporting), methodology (equal), data interpretation (equal), visualization (lead);   
ADB: investigation (equal), data curation (equal), conceptualization (supporting), methodology (supporting), writing - review and editing (supporting);
JMLM: funding acquisition, conceptualization (lead), supervision, methodology (equal), investigation (equal), writing - original draft, writing - review and editing (equal)

\subsection*{Acknowledgments}
This work was supported by the Minerva Foundation, Munich, Germany.  JMLM thanks the Quantum Theory Project at the University of Florida for their hospitality. Computational work was carried out on the Faculty of Chemistry's high-performance computing facility CHEMFARM, which is supported in part by the Ben May Center for Chemical Theory and Computation. GHJ was funded by the National Science Foundation (grant CHE-2430408, ``Advances in Coupled Cluster Theory''; deceased PI: John F. Stanton; current PI: Alberto Perez). 

%

\subsection*{Conflict of interest}

The authors declare no potential conflicts of interest.

\subsection*{Supplementary Material}

\jmlm{Spin-orbital coupled-cluster intermediate equations represented using Brandow (antisymmetrized Goldstone) diagrams;}
Microsoft Excel workbook with the relevant total and interaction energies.

Additional raw data may be obtained from the corresponding author upon reasonable request.

\bibliography{postCCSDTq}



\vspace*{12pt} 
            \begin{figure*}[h!]
            \centering

\subsection*{Table of Contents Graphic ($5\times13$ cm)}

\includegraphics[width=13cm]{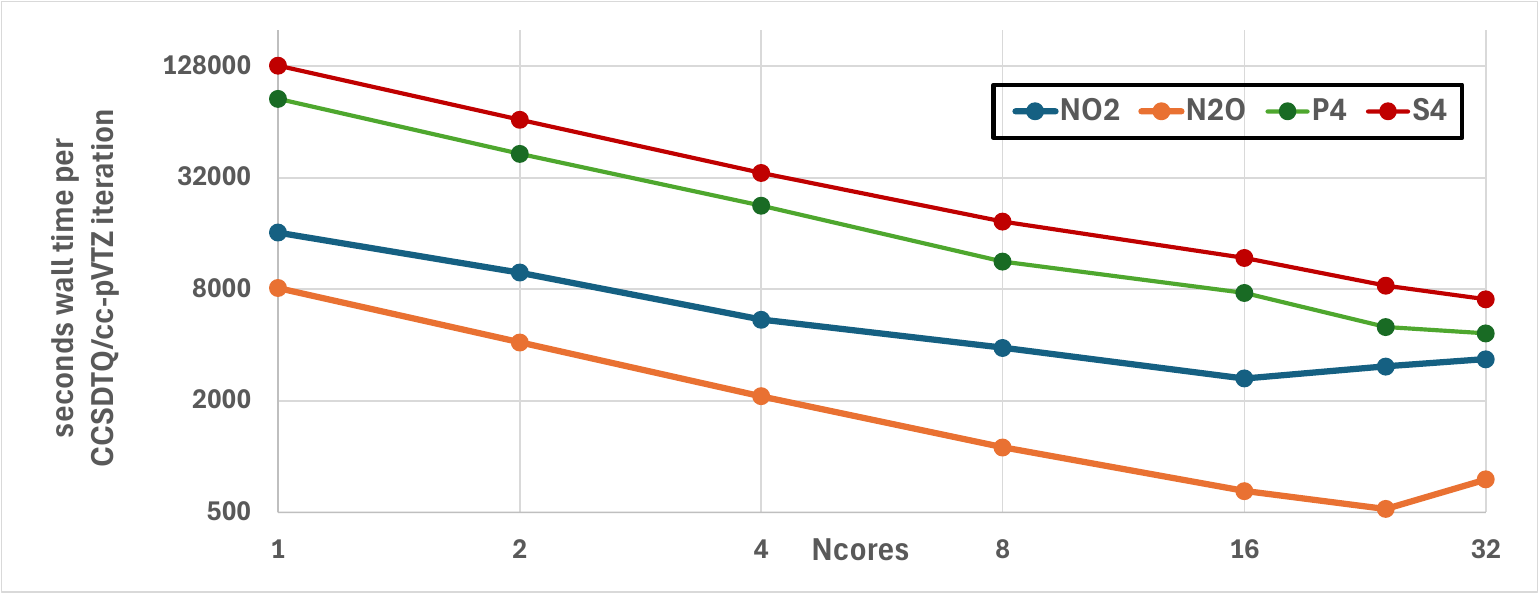}

    \addtocounter{figure}{-1}
\end{figure*}




\clearpage
\onecolumngrid

\setcounter{page}{1}
\setcounter{section}{0}
\setcounter{table}{0}
\setcounter{figure}{0}
\setcounter{equation}{0}
\renewcommand{\thepage}{S\arabic{page}}
\renewcommand{\thesection}{S\arabic{section}}
\renewcommand{\thetable}{S\arabic{table}}
\renewcommand{\thefigure}{S\arabic{figure}}
\renewcommand{\theequation}{S\arabic{equation}}
\renewcommand{\theHsection}{S.\arabic{section}}
\renewcommand{\theHtable}{S.\arabic{table}}
\renewcommand{\theHfigure}{S.\arabic{figure}}
\renewcommand{\theHequation}{S.\arabic{equation}}

\begin{center}
{\large\bfseries Supplementary Information for:\par}
\vspace{0.5em}
{\large\bfseries A new open-shell CCSDTQ implementation and its application to the basis set convergence of post-CCSDT(Q) corrections in computational thermochemistry\par}
\vspace{1.5em}
Aditya Barman$^{\dagger}$, Gregory H. Jones$^{\dagger,*}$, and Jan M. L. Martin$^{*}$

\vspace{0.75em}
{\small $^{\dagger}$Equally contributing authors}

\vspace{0.75em}
CPLETT-26-1042: Revised manuscript \today
\end{center}

\section{A note on dealing with UHF bifurcation}

\begin{table*}

\caption{Total energies (hartree) at different levels of theory, with the cc-pVDZ basis set, for FOO and ClOO with three different reference determinants.\label{tab:bifurcate}}
\begin{tabular}{llllllll}
\hline\hline
                                              & \ce{FO2} low$S^2$ & \ce{FO2} ROHF & \multicolumn{1}{l}{\ce{FO2} high$S^2$} & \ce{ClOO} low$S^2$ & \ce{ClOO} ROHF & \ce{ClOO} high$S^2$ \\
\textless{}$S^2$\textgreater{} & 0.763 & 0.750 & 1.522 & 0.765 & 0.750 & 1.547 \\
\hline
SCF                         & -248.901 650 & -248.894 914 & -248.955 641 & -609.016 259 & -609.009 691 & -609.062 813 \\
MP2                         & -249.475 542 & -249.476 806 & -249.451 165 & -609.527 049 & -609.528 544 & -609.522 658 \\
CCSD                        & -249.490 713 & -249.490 528 & -249.491 488 & -609.554 635 & -609.554 633 & -609.558 341 \\
CCSD(T)                     & -249.515 655 & -249.515 628 & -249.510 228 & -609.578 730 & -609.578 685 & -609.574 472 \\
CCSD(T)$_\Lambda$           & -249.513 762 & -249.513 701 & -249.509 049 & -609.576 627 & -609.576 546 & -609.573 749 \\
CCSDT                       & -249.517 427 & -249.517 379 & -249.517 347 & -609.581 209 & -609.581 170 & -609.581 017 \\
CCSDT(Q)                    & -249.521 754 & N/A          & -249.521 445 & -609.585 246 & N/A          & -609.584 855 \\
CCSDT(Q)$_\Lambda$          & -249.521 754 & -249.521 755 & -249.521 445 & -609.585 246 & -609.585 246 & -609.584 855 \\
CCSDTQ                      & -249.521 485 & -249.521 462 & -249.521 420 & -609.585 142 & -609.585 121 & -609.585 053 \\
\hline\hline
\end{tabular}
\end{table*}

For two of the radical systems in the W4-08 dataset, namely, \ce{FOO} and \ce{ClOO}, bifurcation of the UHF solution has previously been documented by Denis and Ornellas.\cite{Denis2008SpinContaminationXOO} (This is actually a special case of the whimsically named `triplet instability in doublets' described by Stanton and coworkers.\cite{Szalay2004TripletInstabilityDoublets}) In both cases, one solution has an $<S^2>$ near the pure doublet value of 0.75, while the other has a much lower SCF energy and an elevated $<S^2>$ of 1.52 for \ce{FOO} and 1.55 for \ce{ClOO}. We will refer to these solutions as `LowS2' and `HighS2' below.

Convergence to either solution from the default initial guess is somewhat unpredictable in CFOUR; however, the QCSCF converger typically leads to the `HighS2' solution or can be manipulated into doing so. We were able to reliably converge to the `LowS2' solution by the expedient of reading in the density from a converged ROHF calculation as the initial guess.

As a third series of data, we carried out restricted open-shell coupled cluster calculations. In Table~\ref{tab:bifurcate}, one can find results in the cc-pVDZ basis set for all three reference functions at levels through CCSDTQ.

A number of things become immediately apparent. For one, the energetic gap between `HighS2' and ROHF references is quite large at the SCF level (47 m$E_h$ for \ce{ClOO}, 54 m$E_h$ for \ce{FOO}) and still in the several m$E_h$ range at CCSD(T) and CCSD(T)$_\Lambda$ levels, but at the fully iterative CCSDT level plunges down to 192 $\mu E_h$ for \ce{ClOO} and just 80 $\mu E_h$ for \ce{FOO}.

Second, it is very clear that the energies with `LowS2' and `ROHF' references are much closer to each other than either is to `HighS2'. Notably, for \ce{FOO} the gap is just 49 $\mu E_h$ at the CCSDT level, and drops further to 23 $\mu E_h$ at the CCSDTQ level; for \ce{ClOO}, the corresponding numbers are 39 and 21 $\mu E_h$, respectively.

Third, inspection of the CC amplitudes for \ce{FOO} reveals that `HighS2' has a largest $T_1$ amplitude of 0.40 (!) at the CCSDT level, and 0.42 at the CCSDTQ level; the corresponding values are 0.12--0.13 for `LowS2' and 0.11 for `ROHF'. A milder version of the same phenomenon is seen for \ce{ClOO}.

Fourth, while CCSDT(Q) struggles with the `HighS2' -- `LowS2' relative energies (in fact, getting the wrong sign for \ce{FOO}), the powerful CCSDT(Q)$_\Lambda$ method\cite{KallayGauss2005} clearly holds its own, in fact predicting `LowS2' and `ROHF' referenced results to be within 1 microhartree of each other.

\section{Perturbation theory diagrams for intermediates}

These are presented, rotated for readability, on the next three pages.


\clearpage

\begin{figure}[p]
    \centering
    \rotatebox{90}{\begin{minipage}{0.82\textheight}
    \centering
    \includegraphics[width=\linewidth]{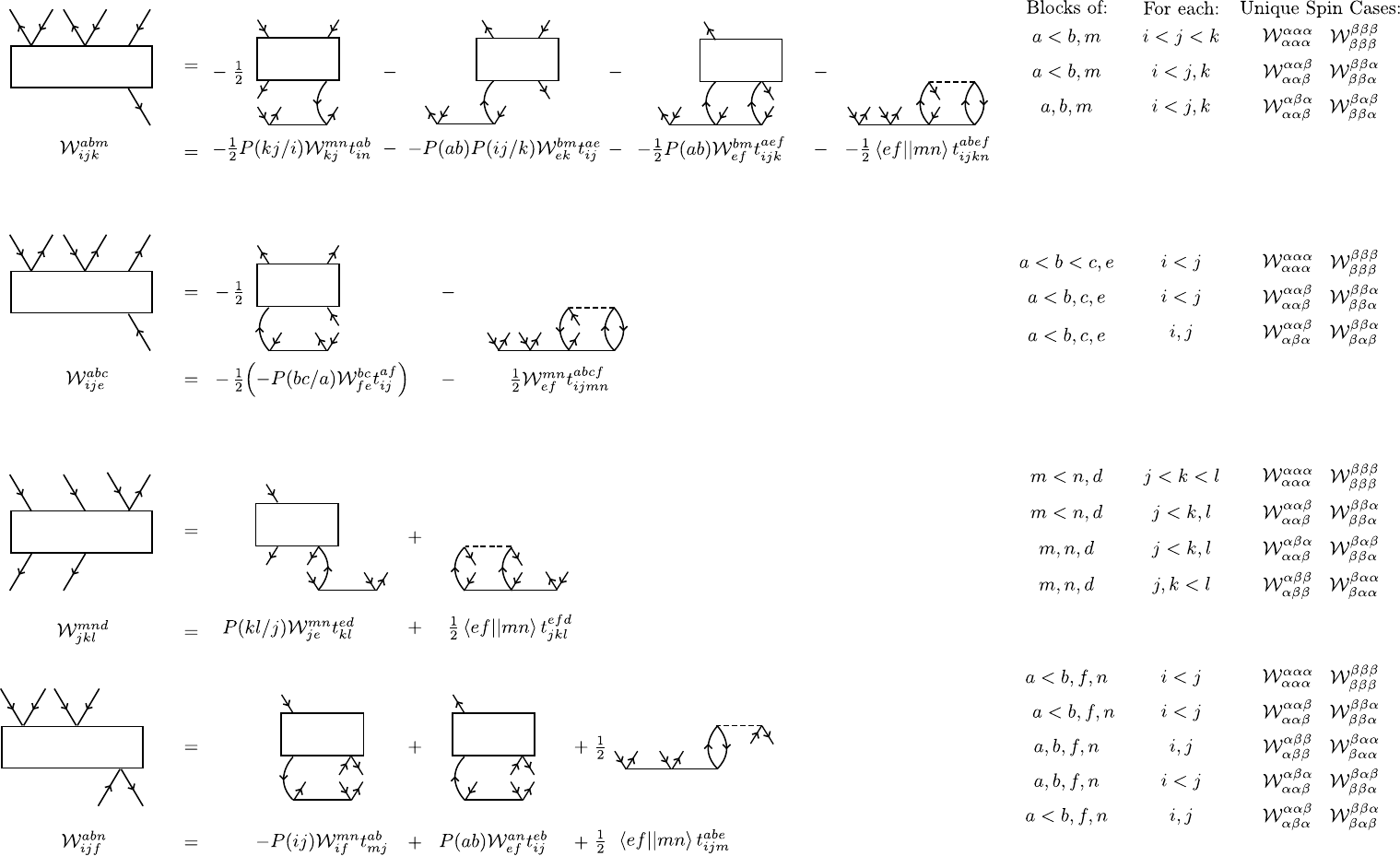}
    \caption{Diagrammatic representation of the intermediates used in the CCSDTQ equations, along with accompanying equations. Einstein summation convention used throughout. $P(pq)$ is defined such that $P(pq) Z(pq) = Z(pq) - Z(qp)$, while $P(pq/r)$ is defined such that $P(pq/r) Z(pqr) = (1 + P(pr) + P(qr)) Z(pqr) = Z(pqr) + Z(qrp) + Z(rpq)$. 4-index $\mathcal{W}$ intermediates correspond to the unprimed $\mathcal{W}$ intermediates from Refs. \cite{gauss_analytic_2000,gauss_analytic_2002}.}
    \label{si:w3}
    \end{minipage}}
\end{figure}
\clearpage

\begin{figure}[p]
    \centering
    \rotatebox{90}{\begin{minipage}{0.82\textheight}
    \centering
    \includegraphics[width=\linewidth]{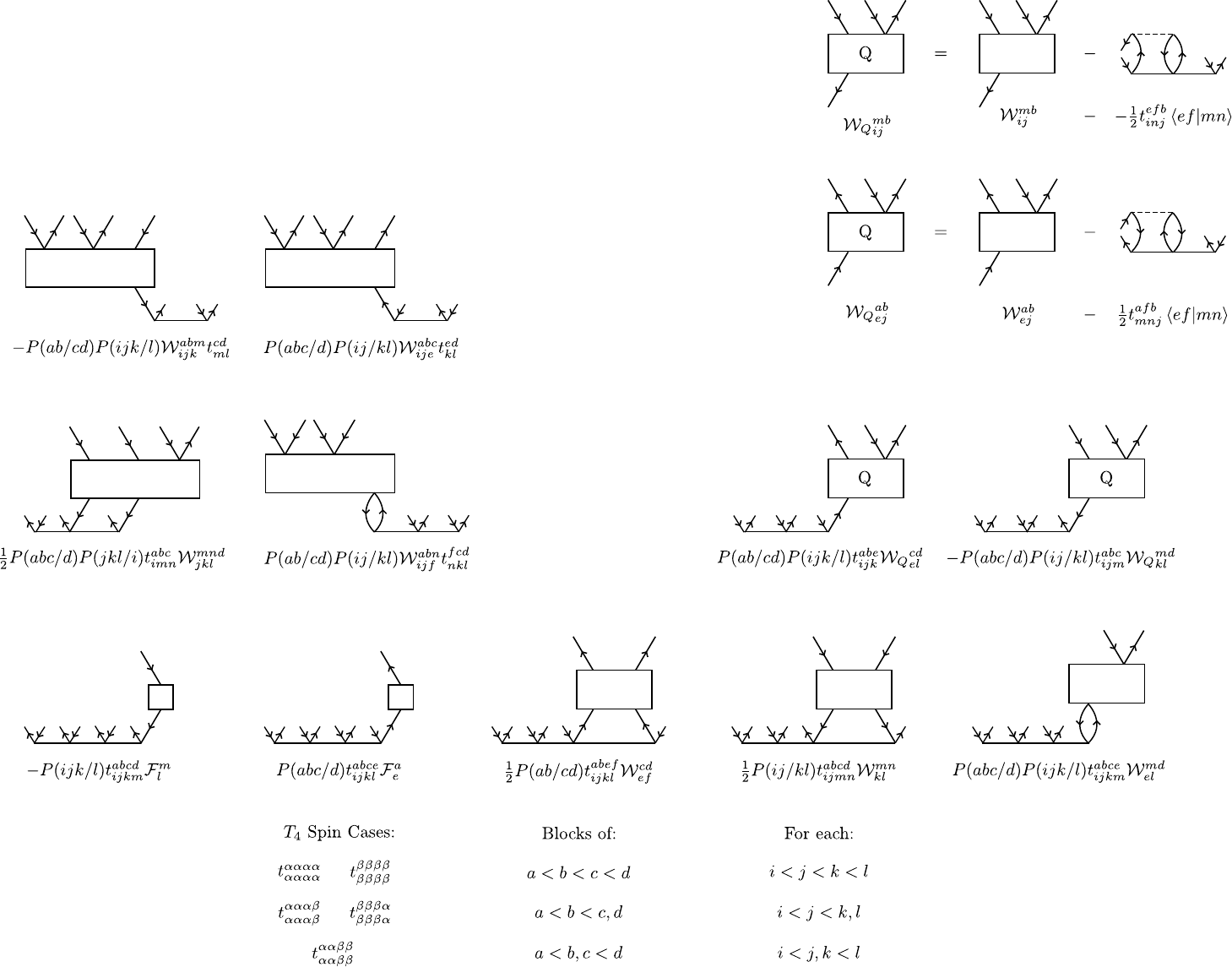}
    \caption{Diagrammatic representation of the contributions to $T_4$ amplitude equations in CCSDTQ with accompanying equations. Einstein summation convention used throughout. $P(pq/rs)$ is defined such that $P(pq/rs) Z(pqrs) = (1 + P(pr) + P(ps) + P(qr) + P(qs) + P(pr)P(qs)) Z(pqrs)$. 4-index $\mathcal{W}$ intermediates correspond to the unprimed $\mathcal{W}$ intermediates from Refs. \cite{gauss_analytic_2000,gauss_analytic_2002}.}
    \label{si:z4}
    \end{minipage}}
\end{figure}
\clearpage

\begin{figure}[p]
    \centering
    \rotatebox{90}{\begin{minipage}{0.82\textheight}
    \centering
    \includegraphics[width=\linewidth]{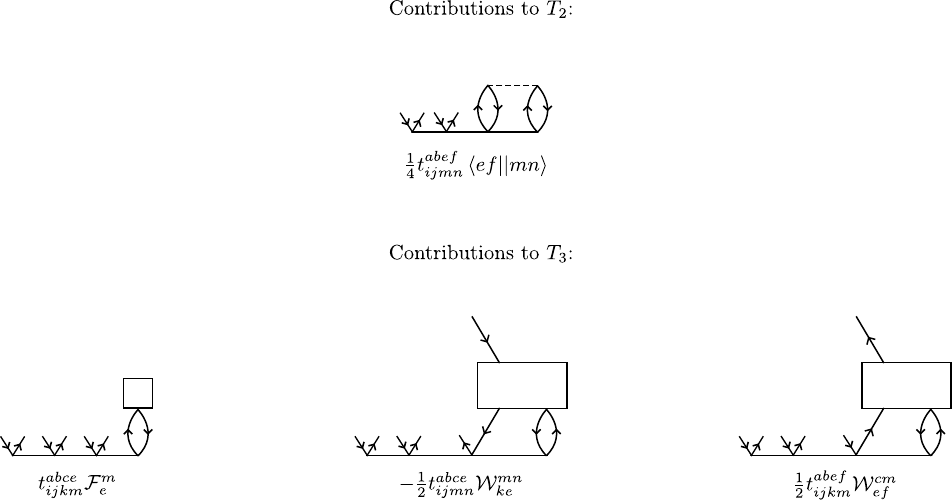}
    \caption{Diagrammatic representation of the contributions to $T_2$ and $T_3$ amplitude equations in CCSDTQ with accompanying equations. 4-index $\mathcal{W}$ intermediates correspond to the unprimed $\mathcal{W}$ intermediates from Refs. \cite{gauss_analytic_2000,gauss_analytic_2002}.}
    \label{si:z23}
    \end{minipage}}
\end{figure}
\clearpage

\end{document}